# Agent Madoff: A Heuristic-Based Negotiation Agent For The Diplomacy Strategy Game

Tan Hao Hao
School of Computer Science and Engineering, Nanyang Technological University, Singapore
tanh0207@e.ntu.edu.sg

*Abstract* – In this paper, we present the strategy of Agent Madoff, which is a heuristic-based negotiation agent competing in the upcoming Automated Negotiating Agents Competition (ANAC 2017). Agent Madoff is implemented to play the game Diplomacy, which is a strategic board game that mimics the situation during World War I. Each player represents a major European power which has to negotiate with other forces and win possession of a majority supply centers on the map. We propose a design architecture which consists of 3 components: heuristic module, acceptance strategy and bidding strategy. The heuristic module, responsible for evaluating which regions on the graph are more worthy, considers the type of region and the number of supply centers adjacent to the region and return a utility value for each region on the map. The acceptance strategy is done on a case-by-case basis according to the type of the order by calculating the acceptance probability using a composite function. The bidding strategy adopts a defensive approach aimed to neutralize attacks and resolve conflict moves with other players to minimize our loss on supply centers.

## 1 INTRODUCTION

Negotiation is described as a process of reaching an agreement between two or more individuals. Technically, we could also treat negotiation as a distributed search through a space of possible agreements. This serves as the fundamental perspective adopted by automated negotiation, which agents devise algorithms to search for Pareto optimal deals within a given negotiation agreement space. Automated negotiation has been a growing area of research in the recent years with an increasing number of applications in domains such as e-commerce, board games and even human-agent negotiation.

The Automated Negotiation Agents Competition (ANAC) is one of the competitions which fuels research interest in developing practical, state-of-the-art agents that can negotiate under various circumstances. As it evolves towards a more practical approach, the latest ANAC 2017 introduces a new negotiation league named as "Negotiation Strategies for the Diplomacy Strategy Game", which requires the participating agents to be able to negotiate over a large agreement space. Under this league, the Diplomacy game almost simulates how human individuals interact during a negotiation process. Hence, this is a much more surreal condition which is closer to the human negotiation environment, and with no doubt the complexity should increase significantly.

This paper first gives an introduction of the above stated league by specifying how the Diplomacy game is played, and how negotiation is done in Diplomacy under the BANDANA framework. Then, it would focus on describing the negotiation strategy used by Agent Madoff, an agent implemented in accordance with the regulations proposed in the above stated league. In addition, the experimental results of the agent's evaluation are provided, and further improvements that could be done on the agent are highlighted.

## 2 MAIN CONTENT

### 2.1 RULES OF ANAC 2017

The upcoming Automated Negotiating Agents Competition (ANAC 2017) has implemented some drastic changes on its rules and regulations compared to the past few years [1]. The league that we are focusing on in this paper is called "Negotiation Strategies for the Diplomacy Strategy Game", which requires participants to implement a negotiation algorithm on top of a ready-made strategic module, combining both modules to form an agent that can play the classical Diplomacy board game.

Unlike the GENIUS framework used by previous years of ANAC, the main differences between Diplomacy Game League are as follow:

- There is no explicit formula to determine an agent's utility function. No graph could be plotted out explicitly like scenarios in the GENIUS framework as before. The assumptions which were made in the study
- Only heuristic approach can be used to estimate the value of a deal proposed and the value of the agent's current utility.
- BANDANA framework does not allow the agents to learn opponent's strategy by now, unlike last year's GENIUS framework

Hence, a totally different paradigm is needed to deal with the BANDANA framework, however common ground between these two frameworks should also be found so that research progress in the GENIUS framework can be further adopted into agents developed in the BANDANA framework.

## 2.2 THE DIPLOMACY GAME

Diplomacy is a strategy game published by Avalon Hill [2] designed for 7 players. Each player represents one of the 7 "Great Powers in Europe" in the years prior to World War I, namely England, France, Germany, Italy, Austria, Turkey and Russia. Each player has a number of armies and fleets positioned on the map of Europe, and the goal is to conquer half of the "Supply Centers" across Europe, (in normal cases, more than 18 supply centers). Other than conquering more than 18 supply centers to win the game (in this case called a "solo victory"), players can also choose to propose draw with other players existing on board to "share" the victory.

Figure 1 below shows a classical map initial setting of the Diplomacy game. The black dots on some of the provinces represents that the province is a supply center.

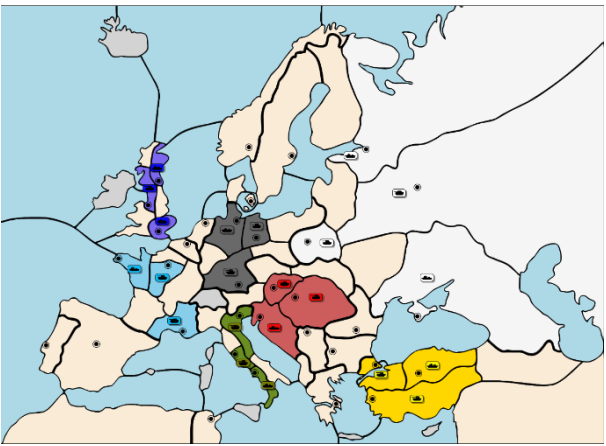

Figure 1: Initial game setting for Diplomacy game

At the start of the game, each powers are allocated with 3 units (except Russia is allocated with 4 units). The game then iterates through 5 types of phases in the following order, starting from year 1901:

- Spring phase
- Summer phase
- Fall phase
- Autumn phase
- Winter phase

Hence the game will develop in the manner of "Spring 1901, Summer 1901, Fall 1901, Winter 1901, Spring 1902 ..." and so on. In each round, all players should submit different types of "orders" for all of their units, depending on which phase they are in.

### 2.2.1  SPRING AND FALL PHASES

For Spring and Fall phases, each player must choose among the following 3 orders to be submitted for each of his unit:

- A "**move to**" order (in BANDANA framework, it is implemented as an MTOOrder class) – where the unit tries to move from one province to another adjacent province.
- A "**hold**" order (implemented as HLDOrder class) – where the unit tries to stay where it is by now
- A "**support move**" order (implemented as SUPMTOOrder) – where the unit does not move, but gives extra strength to another unit that is moving towards the province adjacent to the current unit
- A "**support hold**" order (implemented as SUPOrder) – where the unit does not move, but gives extra strength to another unit to hold its position on the province adjacent to the current unit

If 2 or more units are to move into the same province, the unit which gains more support from other units can successfully move into the intended province. If both units have the same strength, then both units will bounce back to their original position and no one gets in. However, if another unit is moving into the location of the supporting unit, then the support order is "cut", and hence unsuccessful.

### 2.2.2  SUMMER AND AUTUMN PHASES

For Summer and Autumn Phases, there may be some units that are "kicked out" from their original provinces as they are conquered by other units. These units are said to be "dislodged". For such units, they should retreat to another province which is adjacent to its current location (this is also done by the "move to" order). If no such province exists, then the unit must be disbanded, and the player lost one unit.

### 2.2.2  WINTER PHASES

If a player successfully conquers a new supply center throughout the year, then the player is the owner of the supply center. If the player has more supply centers than units, he can build new units (either army unit or fleet unit) at this phase until the number of units equal the number of conquered supply centers. However if the player has more units than supply centers, then some of the units would be disbanded. Players who lost all his units will be eliminated from the game.

Note that army units can only move on inland provinces, and fleet units can only move on sea provinces. There are also coastal provinces which allow to have 1 army unit and 1-2 fleet unit in the province at the same time.

## 2.3 NEGOTIATION IN DIPLOMACY

It is extremely hard for one single player to win the Diplomacy Game without gaining support from the other players, hence negotiation plays a very important role in this game to form alliances among players and agree on certain commitments and promises in order to reach each player's own objective.

Negotiation is done using the BAsic eNvironment for Diplomacy playing Automated Negotiating Agents (BANDANA) framework, which is a Java based framework to allow negotiation capabilities in the Diplomacy game [3]. Under this framework, the main types of deals that would be proposed by players among each other in Diplomacy are:

- An **order commitment**: a proposal to a certain power to impose a certain order on a certain phase. For example: you may need France to support your move from Vienna to Galacia on Spring 1903.

    In BANDANA, it is implemented by a tuple,

    $$oc = (y, \phi, o)$$

    where $y$ is "year", $\phi$ is either "Spring" or "Fall", and $o$ is the order object as specified before.

- A **demilitarized zone (DMZ)**: an agreement among a number of powers to not occupy (or move to) a province on a certain phase. For example: France, Germany and Italy agreed on not to occupy the supply centre in Belgium on Spring 1903.

    In BANDANA, it is implemented by a tuple,

    $$dmz = (y, \phi, A, B)$$

    where $y$ is "year", $\phi$ is either "Spring" or "Fall", $A$ is a list of powers, $B$ is a list of provinces intended to be demilitarized.

Hence, a **BasicDeal** object is implemented to contain a list of order commitments and a list of demilitarized zones, as below:

$$d = \{List<oc>, List<dmz>\}$$

which is to be sent by the agents in the end of each phase to the corresponding powers involved in the deal to accept or reject it.

In ANAC, we are required to only implement the negotiation algorithm of the agent and not its tactical module, as our algorithm will be built on top of another tactical module (the DBrane tactical module) to play the game. Moreover, the same tactical module will be used by other negotiation agents as well. Also in ANAC, promises cannot be broken. This means that if France previously agreed to move from Belgium to Holland, it cannot make another deal to move from Belgium to Picardy. The DBrane tactical module will always obey the deals made, which means the unit in Belgium will indeed move to Holland finally.

As reneging on the deals isn't allowed to be made in this framework, it greatly affects the strategy of the agent since now all agents must obey the deals made. This means that there is no chance for an agent to "cheat" on another agent to gain benefit from it.

## 2.4 LITERATURE REVIEW

As the BANDANA framework is still in its early stage, there are limited existing literatures that fully described a negotiation model which suits the scenario of the Diplomacy game stated above, as a large number of them focus on implementing the tactical part of the agent instead of researching on the negotiation part.

The most relevant literature would be a paper introducing an automated negotiation agent, the Diplominator, where in the paper some existing automated agents are introduced, though a large number of them do not have negotiating capabilities.

### 2.4.1 DUMBBOT

Although DumbBot is not a negotiating bot, it provides reference on deriving heuristics for strategies and tactics. It first assign values to each province which consider factors such as the owner of the province, their strength, the competitiveness of the province and the chances for own units to move there. Then, it implements an algorithm that "averages the board" so that the value of a certain province will also be affected by its adjacent province. The tactics developed by the bot will be based on the values of the provinces.

Although DumbBot's heuristic is only used for tactical purpose, it can still be referenced by negotiation algorithms because negotiation deals can only be made after determining a route plan on how to conquer the supply centers on the board. To determine such route plan, we may need a similar heuristic like DumbBot's.

### 2.4.2 DIPLOMAT

Diplomat uses an economic based view of Diplomacy, which negotiation is used to exchange resources and the results of the trades are focused. A simulated "meeting" is also set up and powers that are involved in the deal undergoes an "auction" to determine the winner that makes the deal eventually. It also implements a part of strategy on deceit, but it would not be applicable to our bot.

Diplomat is said to perform worse than DumbBot, despite its sophisticated design on the negotiation algorithm. However, The Market Based approach and the "auction" approach are still interesting and worth looking into further in future.

### 2.4.3 THE ISRAELI DIPLOMAT

The Israeli Diplomat is a sophisticated Diplomacy player which uses a multi-agent architecture designed based on real life war-time government structures. The 'Prime Minister' acts as the representative and opponents only talk to him. If needed, the Prime Minister will pass the suggested negotiation to another role to gain opinion before replying to the opponent. With such architecture, the task of negotiation could be split up to different departments, each focusing on different criteria.

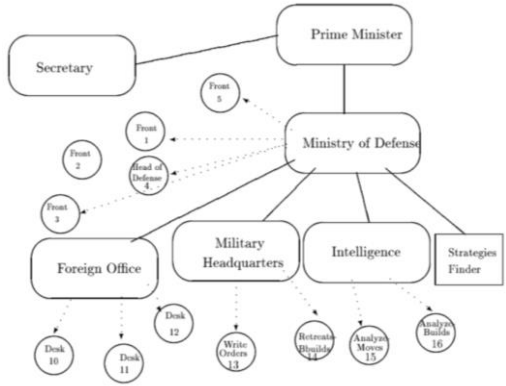

Figure 2: Israeli Diplomat's multi-agent architecture

This multi-agent architecture is claimed to be very successful in the past, and it is truly interesting to be looked into further.

### 2.4.4 DIPLOMINATOR

Diplominator uses a banking protocol as its negotiation strategy, modelling the same idea of how banks decide whether or not to lend money to individuals. The key components of the strategy includes:

- Utility of Cooperation - As a currency is needed in banking protocols, it simply uses DumbBot's heuristic to calculate the destination values of all provinces, and uses it as the currency to measure the values of moves requested by other bots, in order to deduce our utility of cooperating with them.
- Creditworthiness of Opposition Players – A bot should estimate whether the opponent is trustworthy for negotiation or not. A simple way to implement this is to count the percentage of times the opponent has assisted us. Another way will be to calculate the enemy factor, which is defined by

$$enemy\ factor = \frac{\#true\ attacks}{\#chances\ to\ attack}$$

When chances = 0, enemy factor equals the number of true attacks by the player.

- Credit Limit - We should stop assisting opponents at some point until they pay us back, in case we are being too kind to be exploited by opponents in the end.

### 2.4.5 ANACEXAMPLENEGOTIATOR

ANACExampleNegotiator is a random negotiator. It first checks whether there are incoming messages in the message queue, and check what type of message it is. If there are proposed deals coming in, it accepts the deal on a 50% probability. After handling all messages, it picks among 10 randomly generated orders that yields the higher number of supply centers if obeyed by others.

This bot shows a simple two-layered architecture which focus on the acceptance strategy and the bidding strategy, where two strategies may be loosely coupled. However, the framework manual did remind that maximizing the number of supply centers on every phase may not be an optimal strategy.

## 2.5 AGENT STRATEGY

We have designed Agent Madoff, which is a negotiating agent to participate in the upcoming Diplomacy League in ANAC 2017.

The design architecture of Agent Madoff mainly consists of 3 components: heuristic module, acceptance strategy and bidding strategy. The pseudo code of the negotiate() method of Agent Madoff is as below:

```
while (deadline is not reached){
  while(has message){
    handle incoming message;
    if (near to deadline) break;
  }
  bidding strategy;
}
Update hostility and strength;
```

The algorithm enters a while loop which terminates when the deadline is reached. Within the loop, we first handle incoming messages if there are any, and apply the acceptance strategy here. We break the loop if it is near to deadline to prevent opponents from flooding us with incoming messages. Next, we execute our bidding strategy to propose deals to other powers. Before the turn ends, we update the hostility and strength of our opponents to keep track of the game's situation.

In the following discussion, we will further describe our strategy according to the 3 components specified in the design architecture.

### 2.5.1 HEURISTIC MODULE

The heuristic module aids the agent's decision making by evaluating which provinces/regions on the graph are more worthy, and hence have a higher utility for the agent to either move into it or to protect it from opponent's invasion.

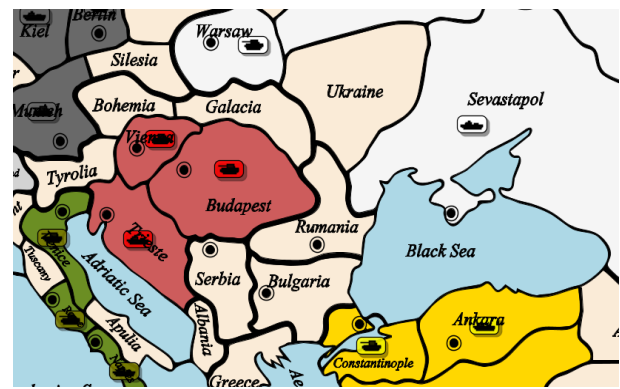

Figure 3: Map representation around Austria (the red region)

Take an example of Austria, certainly supply centers (with a black dot) such as Vienna, Budapest and Trieste have a higher utility for the power itself. However, although provinces like Bohemia, Galacia and Tyrolia are not SCs, they are crucial for Austria because they are adjacent to other SCs which Austria may want to conquer. Hence, the number of adjacent SCs should also be a factor of determining the actual utility of a certain province.

With the idea of adjacent provinces giving influence to its neighbors to result in the final utility value, we propose a heuristic calculation as below:

1. Initialize the utility value for each region: 1 for non-supply centers and 10 for supply centers.
2. Then, we add a value of the sum of utility values of adjacent regions, multiplied by a discount factor of 0.3 to the utility value of each region.
3. We divide the utility value by the maximum utility value among all regions to normalize the value between [0, 1]. This is because in the future calculation of acceptance probability it needs to yield a result between [0, 1], hence all parameters need to be normalized.

An example will be like Figure 3 as below:

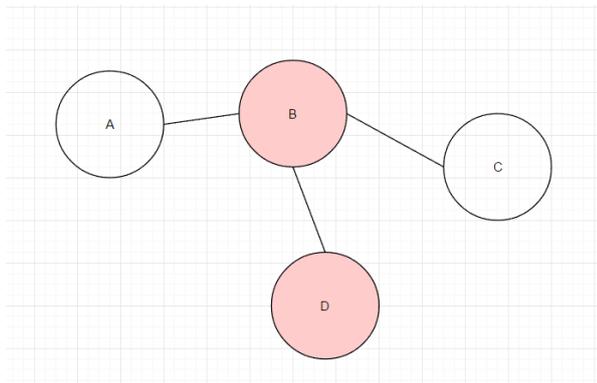

Figure 4: An example of showing the utility calculation

In Figure 4, assume that the regions in red are supply centers. Since B itself is a supply center, it has an initial utility of 10. The sum of the utility value of its adjacent regions are 1+10+1 = 12. Hence the final utility value of B is 10 + 0.3 x 12 = 13.6. Assume that utility value of A = 3, C = 4, D = 15, then the normalized utility value of B is $\frac{13.6}{15}$ = 0.9067.

For our agent, all regions would have a fixed utility value throughout the game. Hence, it could be pre-computed once the agent is initialized.

### 2.5.2 ACCEPTANCE STRATEGY

The acceptance strategy is done on a case-by-case basis according to the type of the order. For each type of order, we determine several parameters which are essential to the acceptance of the order. A composite function that takes in the parameters is derived to calculate a value between [0, 1], and we called it as the acceptance probability for that particular order.

There are 2 types of parameters that will be frequently used in the evaluations below:

**(1) Hostility**

It shows the measure of friendliness of a particular power towards you (e.g. have they attacked you, have they support you before, etc.).

Throughout the game, a hostility list is maintained. For each power, the hostility value is initially set to 0. If a power supports us, its hostility value increase by 5. If a power attacks us (steal our provinces), its hostility value is decreased by 10. The hostility value, h is normalized linearly:

$$If\ h < 0,\ h_{normalised} = 0.5 \left( \frac{h - h_{min}}{-h_{min}} \right)$$

$$If\ h \geq 0,\ h_{normalised} = 0.5 \left( \frac{h}{h_{max}} \right) + 0.5$$

Where $h_{min}$ and $h_{max}$ denotes the minimum and maximum hostility among the list.

**(2) Strength**

It means how strong the particular power is, as we argue that we should team up with the weaker teams to fight the stronger teams so that they will be a balance of power among all players.

The strength of a team is determined by the number of SCs that they have currently conquered. In the beginning each team is given 3 SCs, and with 18 SCs one can declare a solo victory. Hence we need a mapping function to map values from [3, 18] to [0, 1] to represent its strength. In this case, the mapping function used is

$$strength_{normalised} = 0.5 \sin \left( \frac{\pi(\#SCs - 9)}{18} \right) + 0.5$$

**Support move order**

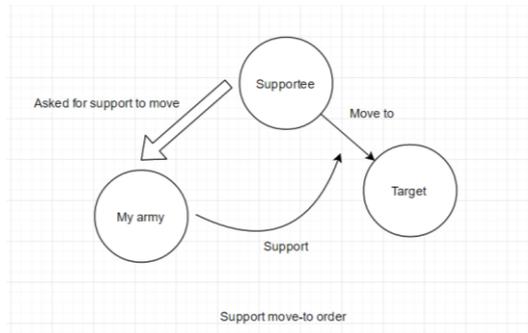

Figure 5: Support move order

We establish 5 parameters: Supportee's hostility, supportee's strength, unit's neediness, target's hostility, target's strength. All parameters have a value between [0, 1].

Unit's neediness, in this case, means how much is the army needed for my team. The higher the unit's neediness, the more reluctant the unit is to support other unit's move. With this, we need to know the original plan of the army which is available in the strategic module of the BANDANA framework:

- If the army is planned to move to some other places, the unit's neediness depends on the utility value of the intended region
- If the army is planned to support my own team, unit's neediness = 1
- If the army is planned to support other teams, its neediness is higher if the supporting team is more friendly towards us
- If the army is planned to just hold, unit's neediness = 0.5

With these 5 parameters, the acceptance probability is then equal to:

$$acceptance\ probability = 0.2 \times hostility_{supportee} + 0.1 \times (1 - strength_{supportee}) + 0.5 \times (1 - unitNeediness) + 0.1 \times strength_{target} + 0.1 \times hostility_{target}$$

If no unit resides in the target region, then target's hostility and strength equals to 0, in this case:

$$acceptance\ probability = 0.3 \times hostility_{supportee} + 0.2 \times (1 - strength_{supportee}) + 0.5 \times (1 - unitNeediness)$$

**Support hold order**

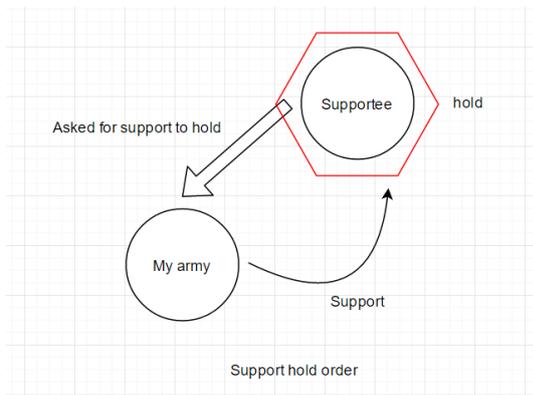

Figure 6: Support hold order

Support hold order works in a similar way with support move-to order without target node, only that the supportee is to hold. Hence,

$$acceptance\ probability = 0.3 \times hostility_{supportee} + 0.2 \times (1 - strength_{supportee}) + 0.5 \times (1 - unitNeediness)$$

**Move-to order**

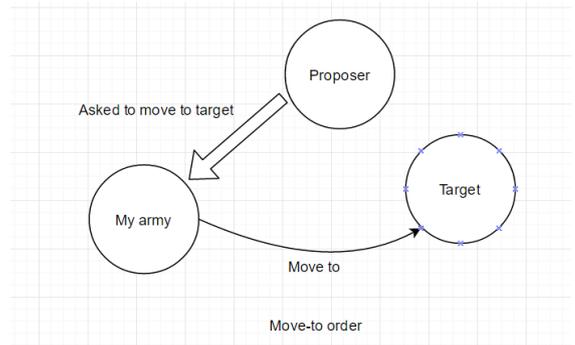

Figure 7: Move-to order

We establish 3 parameters: target's hostility, and whether the new order is better (called *newIsBetter* below).

*newIsBetter* is to determine whether moving to the proposed target is better than its original plan. With this, we check the original plan of the army.

- We initialized *newIsBetter* = 0.2.
- If the army is to hold, and the target to move in has a higher utility value, then *newIsBetter* = 0.8; or else *newIsBetter* remains as 0.2.
- If the army is planning to move to the other destination, we compare the utility of the original destination to the target region. If the target has a higher utility, *newIsBetter* = 0.8; or else *newIsBetter* remains as 0.2.

Hence, the acceptance probability is as below:

$$acceptance\ probability = 0.3 \times hostility_{target} + 0.7 \times newIsBetter$$

If no unit resides in the target region, then target's hostility equals to 0, in this case the acceptance probability simply equals to *newIsBetter*.

**Hold order**

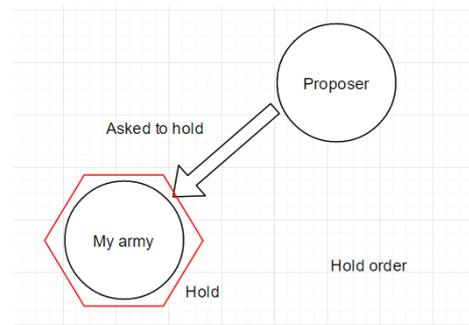

Figure 8: Hold order

We first check the unit's original plan.

- If its plan is to hold, simply accept the proposal.
- If its plan is to move to another region, check the utility of the intended region. If the province has a utility greater than 0.7, reject the proposal. Or else, return a default probability of 0.4. This shows a slight reluctance to accept a hold order for our agent.

- If its plan is to support other units, it is highly improbable to accept a hold order. Hence we return a probability of 0.1.

**DMZ order (demilitarized zone order)**

DMZ proposal is made between several powers such that all the powers agree not to invade the specified provinces.

We establish 2 parameters needed: utility of the region, and the competitiveness of the DMZ proposal.

The higher the utility of the region, the more reluctant we would demilitarize it since it is important to us.

Competitiveness is measured by counting the number of powers involved in the DMZ. It is smoothen by a logarithm function to map the value between [0, 1]. We argue that the more number of powers involved, the harder it is for you to fight for the region if you do not accept the DMZ, hence the more competitive it is the more probable we should demilitarize the region.

For DMZs, we check the original plan of each units and see which provinces we intend to move to.

- If the DMZ zone is not within our intended provinces, simply accept it as it cause no harm to us
- Else, we compute the acceptance probability for the DMZ proposal as below:

$$acceptance\ probability = 0.6 \times competitiveness + 0.4 \times (1 - utilityOfRegion)$$

**COMPOSITE ORDER PROPOSAL**

Proposals usually contain more than 1 type of order. For example, A could send a proposal to B such that A request B to support his move, and A will DMZ one of the regions in conflict with B as a return of favor. In this case, this proposal contains a support order and a DMZ order.

For our agent, after calculating the acceptance probability of each order, we calculate the mean probability of all orders and treat it as the acceptance probability of the composite order proposal. Denote mean probability as $x$:

- If $x > 0.8$, accept the deal;
- If $x < 0.4$; reject the deal;
- If $0.4 \leq x \leq 0.8$, we flip a coin so that the deal is accepted with a probability of $x$.

In addition, our agent only takes in account the proposals regarding the current phase only. For proposals regarding the future phases, we would return an acceptance probability of 0, as we argue that there is no reason for us to bind to a "virtual" agreement made for the unpredictable future.

**2.5.4 BIDDING STRATEGY**

Agent Madoff does not speculate moves further than the current phase. Hence, each proposal of the agent is done only according to the current phase and game setting. This is because binding to deals which happen in the future may be invalid and hence unnecessary.

The bidding strategy that our agent adopted is a defensive, de-conflict strategy. It consists of 2 stages: neutralize the attack regions, and resolve the conflict regions.

**Order Calculator**

We first discuss about the "order calculator" which is used within the bidding strategy. By assuming that for each type of order, the opponent considers same parameters as our agent, we could use our acceptance strategy module to mimic an order calculator which calculates the acceptance probability of a power for a given order in its own perspective. With this, we could propose orders with higher acceptance probability to achieve a higher chance of agreement.

**Neutralizing attack regions**

The strategic module within BANDANA is able to return the opponent's move given its power and the current game state, before negotiation is made. Hence for each power, we could speculate the pre-determined move of each of its unit. If we find that the unit's move is to invade our supply center, we should neutralize the attack through negotiation.

The neutralization steps are as below:

- We generate all possible orders for the attacking unit, and calculate all acceptance probability of the orders.
- If there exists orders which has a higher probability than the attacking order, we pick the highest one and propose it to the opponent.
- If no alternative orders are found, we request for support from other powers, and propose a "favor returning" deal (which will be discussed later) to increase the chance of agreement.
- If we have no units to move into / hold our supply centers, we could only ask for a DMZ with the attacking unit.

**Resolving conflict regions**

Since the strategic module is able to anticipate the opponents move given the opponent's power and the current game state, we could then see if our agent's moves are in conflict with other agent's move before negotiation takes place.

A "conflict" means that there are more than 1 opponent that want to move to a region that you intend to move into. If there is, then negotiation proposals are needed to resolve the conflict.

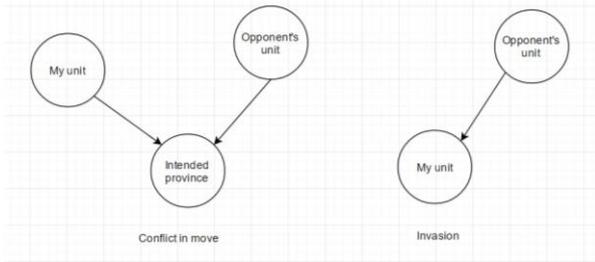

Figure 9: Conflict move

The resolving steps are as below:

- For each power, we check if we have any in-conflict moves with them.
- We generate all possible orders for the opponent's conflicting unit, and calculate all acceptance probability of the orders.
- If there exists orders which has a higher probability than the conflict order, we pick the highest one and propose it to the opponent.
- If no alternatives are found, we request for support from other powers, and propose a "favor returning" deal (which will be discussed later) to increase the chance of agreement.
- We keep a DMZ deal with the opponent's unit for that region as a "reservation deal" which will be proposed should the previous deals failed to be accepted.

**Return the favor**

Normally within a proposal, we would add a "favor returning" deal apart from a pure request of support so that the proposal is more likely to be accepted. In this case, we specify a "return credit", which is the number of our units which could return the favor to the powers which have supported us. This is to prevent most of our units being used for returning the favor, which further weakens our strength. We allow 1/3 of the total number of units that we currently have to be our "return credit".

We propose 2 ways to return the favor to our supporters:

- By supporting its move: we search for units situated adjacent to the power's army's adjacent neighbors, as shown in the top part of the diagram above. If there is, and if the neighbor (region A as in Figure 9) has a very high utility for the power, we could propose to support the power to move into A in return.

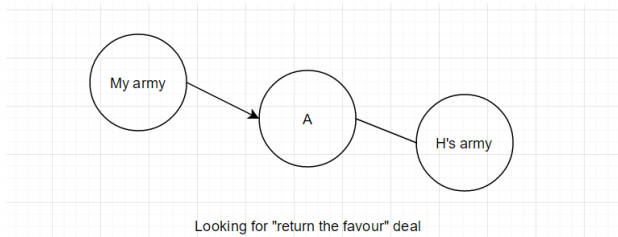

Figure 10: Return the favor by support move-to order

- By supporting it to hold: some units may want to hold their position to prevent invasion into their regions by other powers. If we have a unit which is adjacent to the supporter's unit, we could support them to hold.

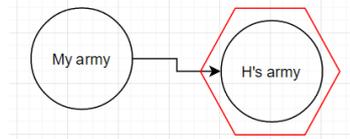

Figure 11: Return the favor by support hold order

## 3 CONCLUSION

In this paper, we first introduce the new Diplomacy league in ANAC and observe how negotiation is implemented in the league using the BANDANA framework. We give a review on existing Diplomacy agents, which provide useful ideas for us to implement an efficient agent. We present the design of Agent Madoff, whose architecture comprises of 3 main components: the heuristic module, acceptance strategy and bidding strategy. Our agent is submitted to participate in ANAC 2017 Diplomacy league, and the results are forthcoming.

Future work will focus on improving each of the modules: for the heuristic module, we could adopt a dynamic model which changes according to the game situation instead of the current static model; for the acceptance strategy, the parameters could be adjusted to further mimic the thought process of a human agent making negotiation decisions.; for the bidding strategy, it could be improved to search for tempting bids that are more likely to be accepted by the opponents. Instead of adopting a defensive strategy, we could also propose deals to form attacks and arrange battle plans against our opponents, acting as a real diplomat during war.

## ACKNOWLEDGEMENT

I would like to thank my supervisor, Prof Bo An for providing detailed guidance and support for this project. His experienced thoughts in the field of automated negotiation has spurred exciting discussions between us. Moreover throughout the year, he has showed me by himself all the essential qualities of being a dedicated and successful researcher.